\documentclass[%
 reprint,
preprintnumbers,
 amsmath,amssymb,
 aps,
]{revtex4-1}

\usepackage{graphicx}
\usepackage{dcolumn}
\usepackage{bm}
\usepackage{epstopdf}
\usepackage{wrapfig,epsfig}

\begin{document}

\preprint{J-PARC-TH-0087}

\title{QCD mechanisms for accessing the nucleon GPDs\\ with the exclusive pion-induced Drell-Yan process at J-PARC}

\author{Kazuhiro Tanaka}
 \email{kztanaka@juntendo.ac.jp}
\affiliation{%
 Department of Physics, Juntendo University, Inzai,
  Chiba 270-1695, Japan}
\affiliation{ J-PARC Branch, KEK Theory Center, Institute of
  Particle and Nuclear Studies, KEK, 203-1, Shirakata, Tokai, Ibaraki,
  319-1106, Japan
}%




\date{\today}

\begin{abstract}
Generalized parton distributions (GPDs) encoding multidimensional information of
hadron partonic structure appear as the building blocks in a factorized description of
hard exclusive reactions. The nucleon GPDs have been accessed by deeply virtual
Compton scattering and deeply virtual meson production with lepton beam. A
complementary probe with hadron beam is the exclusive pion-induced Drell-Yan process.
We discuss recent theoretical advances on describing this process in terms of the
partonic subprocess convoluted with the nucleon GPDs and the pion distribution
amplitudes. Furthermore, we mention the feasibility study for measuring the exclusive
pion-induced Drell-Yan process, $\pi^- p \to \mu^+ \mu^- n$, via a spectrometer at the
High Momentum Beamline being constructed at J-PARC in Japan. We also point out
the possible soft partonic mechanisms beyond the QCD factorization framework, which could
give important contributions at J-PARC kinematics,
and present an estimate of the soft mechanisms making use of dispersion
relations and quark-hadron duality. 
Realization of the measurement of the exclusive pion-induced Drell-Yan process at J-PARC will provide a new test of QCD descriptions of a novel class of hard exclusive reactions. It will also offer the possibility of experimentally accessing nucleon GPDs at large timelike virtuality. 
\end{abstract}

\maketitle


\section{Introduction}
\label{sec:1}
We consider the pion-induced dimuon production. Summing the absolute square of the corresponding amplitudes, 
$\pi N\to q\bar q X \to\gamma^*  X\to \mu^+\mu^- X$, over the accompanying hadronic final state $X$, 
we obtain the inclusive Drell-Yan cross section. The leading contribution comes from the transversely-polarized virtual photon $\gamma^*$, as a consequence of the helicity conservation in the annihilation of the on-shall, massless quark and antiquark associated with the relevant partonic subprocess $q \bar q \to \gamma^*$.
(See e.g. \cite{Peng:2015spa} and references therein.) 

\begin{figure}[b]
{\includegraphics[width=0.33\textwidth]{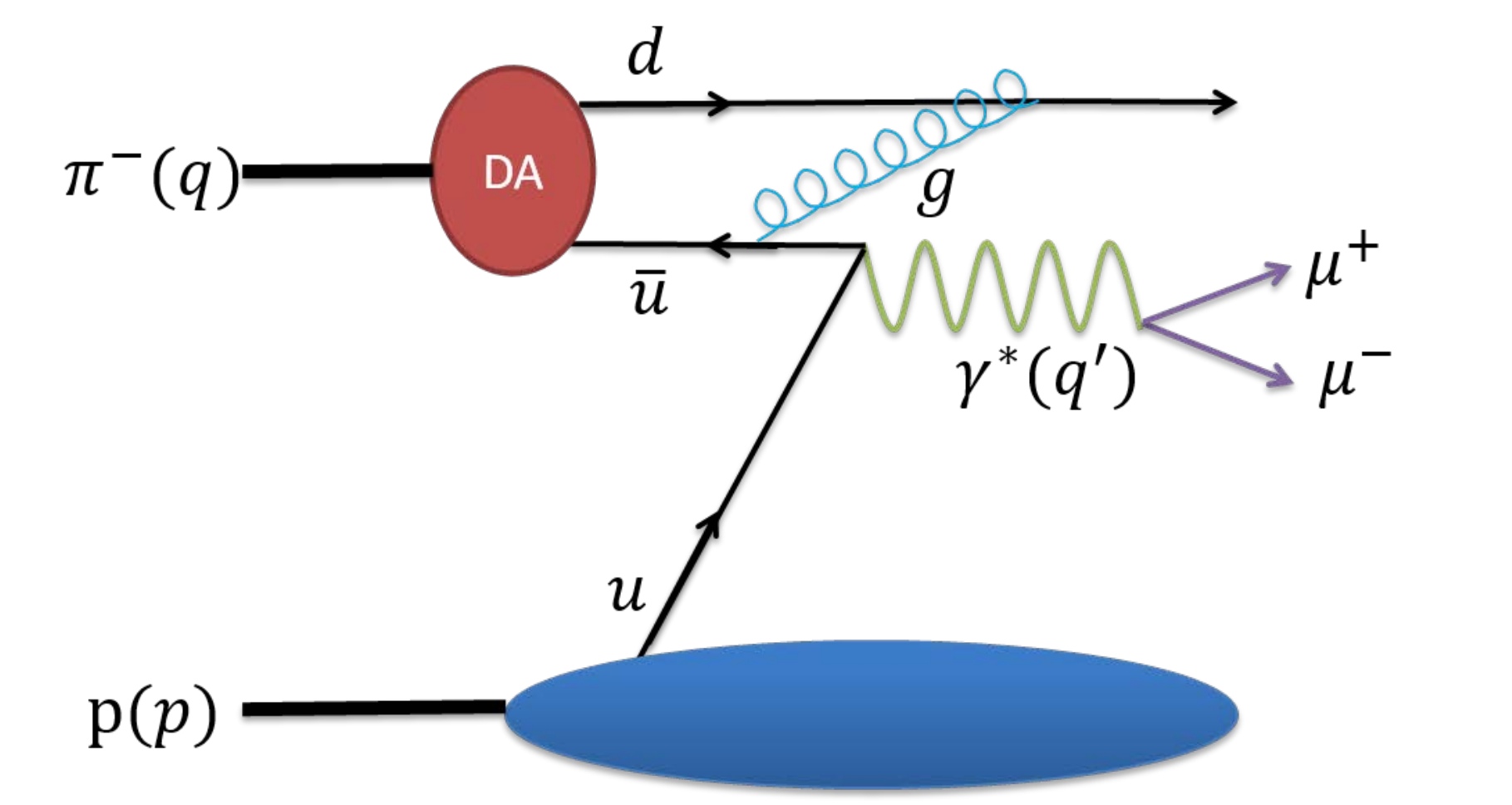}
\label{fig:edycross1}}
\\ \vspace{0.7cm}
{\includegraphics[width=0.33\textwidth]{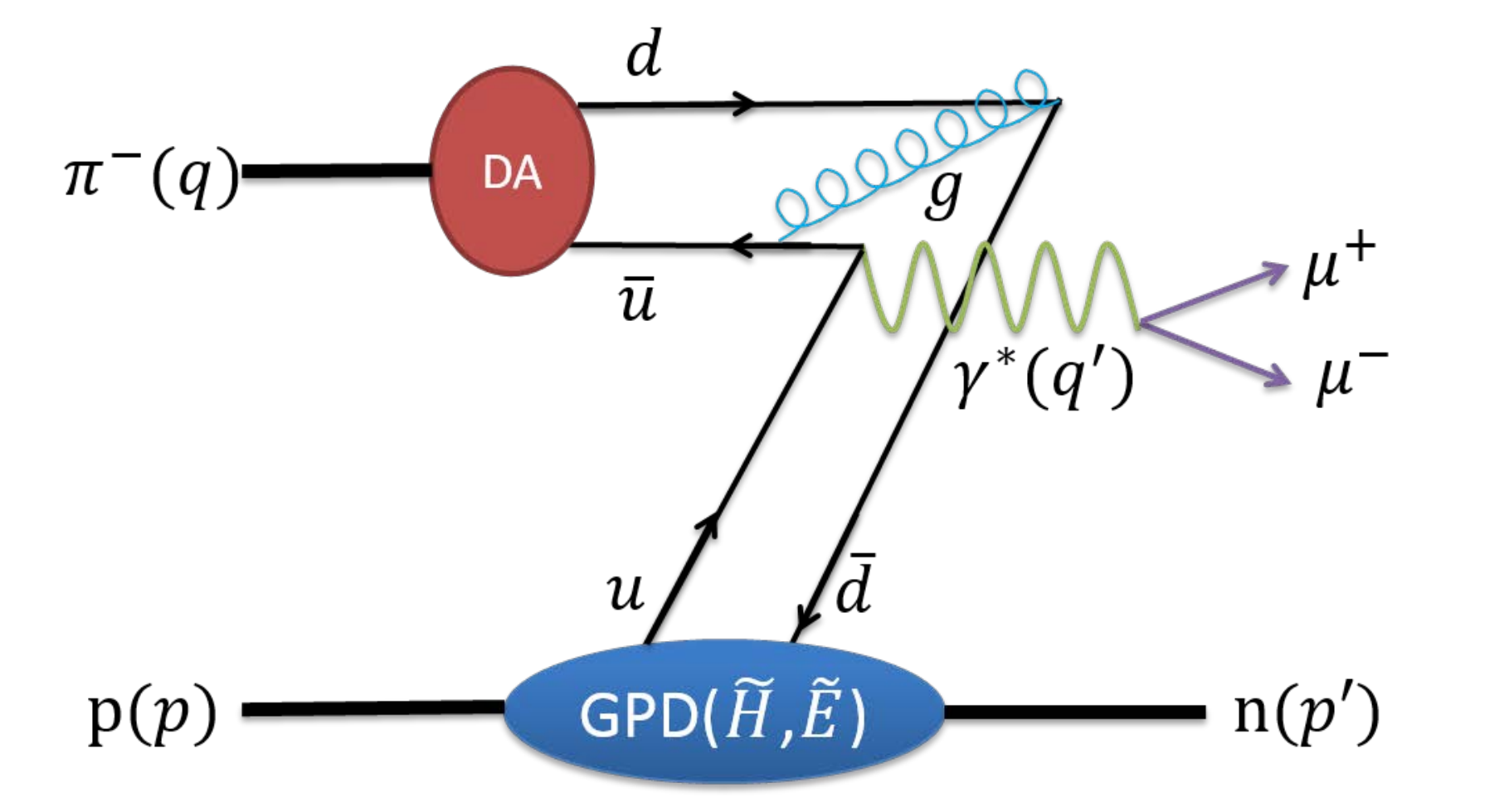}
\label{fig:edycross2}}
\caption{\label{fig1} Inclusive pion-induced Drell-Yan process for the forward
dimuon production (upper)  and exclusive pion-induced Drell-Yan process (lower).}
\end{figure}

On the other hand, the dimuon angular distribution for the production in the forward region is known to obey the pattern associated with the longitudinally-polarized virtual photon, which can be produced 
by the annihilation of off-shell quark $q$ or antiquark $\bar q$.
The relevant off-shellness for quark and antiquark may be 
caused by perturbative gluon exchange of the type in the upper diagram of 
Fig.~\ref{fig1}. Thus, this type of mechanism with the gluon exchange plays important role for the forward production. 
Now, the spectator quark originating from the pion may be absorbed by the target nucleon, 
giving rise to the exclusive final state, $\mu^+\mu^- N$, as represented by the lower diagram in Fig.~\ref{fig1}. 
This is the exclusive Drell-Yan process and this type of diagrams gives important contributions 
for the dimuon production in the forward region
of the exclusive Drell-Yan process~\cite{Berger:2001zn}.

To realize experiments of the exclusive Drell-Yan process, 
we need pion beam with high intensity and moderately high energy, 
such that the cross section 
for each exclusive channel is not strongly suppressed.
With the 
High Momentum Beamline being constructed at J-PARC,
the secondary pion beam with the energies$\sim 15$-$20$~GeV allows such situation.
Thus, the secondary pion beam at J-PARC is best suited to study meson-induced hard exclusive processes 
like exclusive Drell-Yan process.

We discuss the exclusive Drell-Yan process. 
In Sec.~\ref{sec:2}
we discuss a recent estimate of the cross section of the exclusive Drell-Yan process based on the QCD factorization formula, 
represented by the lower diagram in Fig.~\ref{fig1} and the similar diagrams at the same order in $\alpha_s$, as a report of our recent paper
with T.~Sawada, W.~C.~Chang, S.~Kumano, J.~C.~Peng, and S.~Sawada~\cite{Sawada:2016mao}. 
Here we mention the feasibility study for measuring the exclusive
pion-induced Drell-Yan process with the E50 spectrometer at J-PARC.
In Sec.~\ref{sec:3}, we argue that the non-factorizable mechanism, i.e., the mechanism beyond the QCD factorization, 
could play important roles in the exclusive Drell-Yan process and give its first estimate 
based on the light-cone QCD sum rule approach.
We conclude the paper in
Sec.~\ref{sec:summary}.

\section{QCD factorization at LO}
\label{sec:2}
We consider the exclusive Drell-Yan production, $\pi^- p \to \gamma^* n \to \mu^+ \mu^- n$, 
in particular, with
the production of $\gamma^*$ in the forward region corresponding to the small invariant-momentum-transfer,
$t =\Delta^2$ with $\Delta\equiv q-q'$, where $q$ and $q'$ are the momenta of the initial pion and the produced $\gamma^*$. 
In this case, as already demonstrated 
in Sec.~\ref{sec:1}, 
the complete annihilation of quark as well as antiquark from the pion,
as in the lower diagram of Fig.~\ref{fig1},
plays important role. As a result, the relevant amplitude is expressed as the convolution of the corresponding 
partonic (short distance) annihilation processes with the two separate parts of long-distance nature, associated with the pion and the nucleon, respectively:
the upper long-distance part denotes the pion distribution amplitude (DA), whose information can be obtained from, e.g., $\gamma \gamma^* \to \pi^0$ process at Belle and Babar.
On the other hand, the lower long-distance part denotes the generalized parton distribution functions (GPDs) as an 
off-forward nucleon matrix element, whose forward limit reduces to the usual helicity distribution, $\Delta q(x)$.

Here, the GPDs are defined
as the off-forward matrix element of the bilocal light-cone operators of the type, 
$\langle n(p')|\bar q(-y/2)\cdots q(y/2)|p(p)\rangle$, with $y^2=0$,  and are the functions of
the relevant invariants, the (average) 
light-cone momentum fraction $x$ and the skewness $\xi$ ($=(p-p\,')^+/(p+p\,')^+$), 
as well as $t$ (see e.g., Sec.~II in \cite{Sawada:2016mao}).   
Such definition corresponds to merging the nucleon form factor, as the off-forward matrix element of a local operator to describe the spatial distribution of the constituents, and the usual parton distribution function (PDF), as the forward matrix element of a bilocal light-cone operator to describe the deep configuration of the constituents.
Decomposing the above type of matrix elements into the independent Lorentz structures as
($P\equiv (p+p')/2$), 
\begin{eqnarray}
&&2  P^+ \int  \frac{d y^-}{4\pi}e^{i x P^+ y^-}
 \left\langle p' \left| 
 \bar{q}(-y/2) \gamma^+ q(y/2) 
 \right| p \right\rangle \Big |_{y^+ = \vec y_\perp =0}
\nonumber \\
 &&=  \bar{u} (p') 
 \left [ H^q (x,\xi,t) \gamma^+
     + E^q (x,\xi,t)  \frac{i \sigma^{+ \alpha} \Delta_\alpha}{2 \, m_N}
 \right ] u (p)\ ,
\label{eqn:gpd-n}
\end{eqnarray}
and
\begin{eqnarray}
&& 2  P^+\int  \frac{d y^-}{4\pi}e^{i x P^+ y^-}
 \left\langle p' \left| 
 \bar{q}(-y/2) \gamma^+ \gamma_5 q(y/2) 
 \right| p \right\rangle \Big |_{y^+ = \vec y_\perp =0}
\nonumber \\
&& = \bar{u} (p') 
 \left [ \tilde{H}^q (x,\xi,t) \gamma^+ \gamma_5
     + \tilde{E}^q (x,\xi,t)  \frac{\gamma_5 \Delta^+}{2 \, m_N}
 \right ] u (p)\ ,
\label{eqn:gpd-p}
\end{eqnarray}
we obtain the familiar proton GPDs, $H^q, E^q$, for each quark flavor $q$, and also $\tilde H^q, \tilde E^q$ for the case with the additional $\gamma_5$;
here, 
$ \left| k \right\rangle \equiv \left| p (k)\right\rangle$,
$u(k)$ denotes the proton
spinor with momentum $k$ and mass $m_N$, 
and we do not show the gauge-link operator between
two quark fields.
The Ji sum rule for the quark's angular momentum contribution, $J^q = \int_{-1}^{1} dx \, x \, [H^q (x,\xi,0) +E^q (x,\xi,0) ] / 2$, 
demonstrates that the GPDs carry the informations beyond the ``addition'' of 
informations carried by the original two quantities, the form factors and the PDFs.

Those GPDs, $H, E, \tilde H$, and $\tilde E$ are measured by the deeply virtual Compton scattering (DVCS) 
corresponding to $\gamma^* p \to \gamma p'$ process in the experiments at JLab, HERMES, COMPASS, etc., 
and also by the deeply virtual meson production (DVMP).
For the case of the deeply virtual pion production, $\gamma^* p \to \pi p'$,
the pseudoscalar nature of the pion allows us to probe the GPDs $\tilde H^q$ and $\tilde E^q$ solely, 
associated with $\gamma_5$ as in (\ref{eqn:gpd-p}).
Interchanging the initial 
$\gamma^*$ and the final pion in the deeply virtual pion production with the $\gamma^*$ made timelike, 
we obtain the exclusive Drell-Yan process.
This demonstrates that the exclusive Drell-Yan process at the J-PARC allows us to probe $\tilde H^q$ and $\tilde E^q$ solely and plays complementary role compared with the deeply virtual pion production at, e.g., JLab. The kinematical region accessible by the exclusive Drell-Yan process at the J-PARC is also complementary to those accessible by the GPD measurements 
by the various other experiments~\cite{Sawada:2016mao}.

As noted above, the lower diagram of Fig.~\ref{fig1} and similar diagrams at the same order in $\alpha_s$ obey the QCD factorization.
In these diagrams, the ``hard'' gluon exchange ensures that the vertices in the partonic subprocess are separated by short distances, so that they indeed represent the short-distance partonic subprocess combined with the two long-distance parts, the pion DA and the nucleon GPD. Thus, the diagrams associated with this type of gluon exchange give the leading order (LO) in the  factorization formula for exclusive Drell-Yan process.

The first estimate of the exclusive Drell-Yan cross section using the LO factorization formula was performed 
by Berger, Diehl and Pire~\cite{Berger:2001zn}. 
The corresponding cross section at the large $Q'^2\equiv q'^2$ scaling limit
($\tau \equiv Q'^2/(2 p\cdot q)$, $f_\pi$ is the pion decay constant, and $\tilde x=-\xi$),
\begin{eqnarray}
&&\frac{d\sigma_L}{dt dQ'^2}
= \frac{4\pi \alpha_{\rm em}^2}{27}\frac{\tau^2}{Q'^8} f_\pi^2\, \Bigl[ (1-\xi^2) |\tilde{\cal H}^{du}(\tilde{x},\xi,t)|^2
- 2 \xi^2  \nonumber \\
&&\!\!\!\!
\times{\rm Re} \bigl( \tilde{\cal H}^{du}(\tilde{x},\xi,t)^* \tilde{\cal E}^{du}(\tilde{x},\xi,t) \bigr)
   -   \frac{\xi^2t}{4 m_N^2}|\tilde{\cal E}^{du}(\tilde{x},\xi,t)|^2 \Bigr]\ ,
\label{eq_dcross}
\end{eqnarray}
is expressed by the three types of terms:
we have the absolute square of ($e_q$ is the quark's electric charge)
\begin{eqnarray}
\lefteqn{\tilde{\cal H}^{du}(\tilde{x},\xi,t)=\frac{8\alpha_s}{3} \int_{-1}^1
  dz\, \frac{\phi_\pi(z)}{1-z^2} \int_{-1}^1 dx
  \Bigl( \frac{e_d}{\tilde{x}-x- i\epsilon}} 
\nonumber \\ 
 && \;\;\;\;\; \;\;\;\;\; \;\;\;\;\;
- \frac{e_u}{\tilde{x}+x- i\epsilon}
  \Bigr)  \bigl[ \tilde{H}^{d}(x,\xi,t) - \tilde{H}^{u}(x,\xi,t)
  \bigr]\ ,
\label{eq_Hdu}
\end{eqnarray}
i.e., the 
convolution of the hard part, the pion DA $\phi_\pi(z)$ of leading twist, and the GPD $\tilde H^q$,
while the similar convolution $\tilde{\cal E}^{du}(\tilde{x},\xi,t)$ 
with $\tilde H^q$ replaced by the GPD $\tilde E^q$ arises as its absolute square and its interference with $\tilde{\cal H}^{du}$.
In (\ref{eq_Hdu}), the $p\to n$ transition GPDs relevent to exclusive Drell-Yan process of Fig.~\ref{fig1} 
is expressed by the proton GPDs of (\ref{eqn:gpd-p}) using isospin invariance relations 
(see \cite{Berger:2001zn,Sawada:2016mao}). 
The subscript ``$L$'' in (\ref{eq_dcross}) indicates that this cross section 
is obtained at 
the leading twist, associated with 
the production of the longitudinally-polarized $\gamma^*$.
(For a treatment of higher-twist effects allowing the production of the transversely-polarized 
$\gamma^*$, 
see \cite{Goloskokov:2015zsa}.)

\begin{figure}[hbtp]
\includegraphics[width=0.45\textwidth]{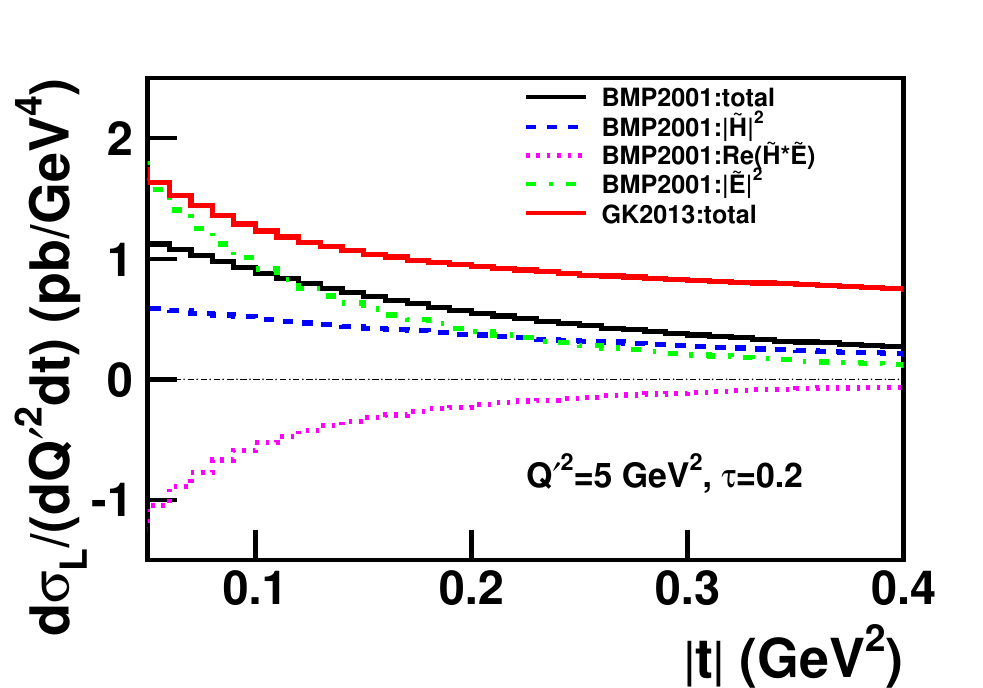}
\includegraphics[width=0.45\textwidth]{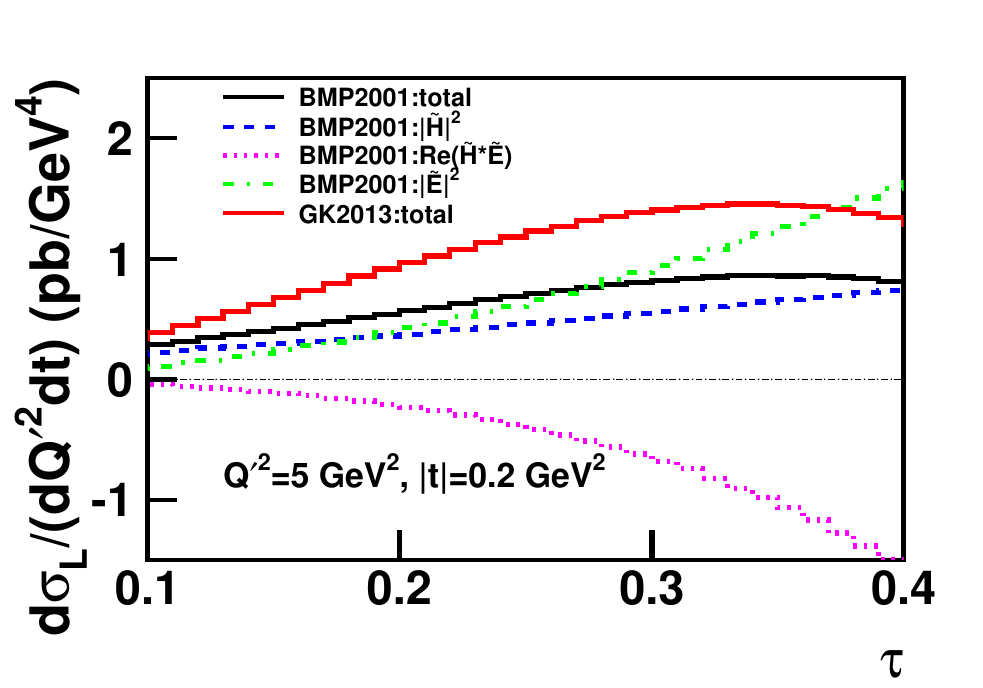}
\caption
[\protect{}] {Differential cross section
of
$\pi^- p \to \gamma^* n$ with $Q'^2=5$~GeV$^2$
as a function of $|t|$ for $\tau = 0.2$ (upper) and as a function of
  $\tau$ for $|t|= 0.2$~GeV$^2$ (lower)~\cite{Sawada:2016mao}. 
The total~(\ref{eq_dcross}) is shown by the black solid line, 
with the individual contributions being shown
for the terms with $|\tilde{\cal H}^{du}|^2$ (dashed), $\mbox{Re} (
\tilde{\cal H}^{du*}\, \tilde{\cal E}^{du} )$ (dotted), and $|\tilde{\cal
    E}^{du}|^2$ (dash-dotted), using the BMP2001 input, while the red solid line
     shows (\ref{eq_dcross}) using the GK2013 input.
    }
\label{fig:edycrossPLB}
\end{figure}
With $Q'^2=5$~GeV$^2$ for the mass of the produced dimuon, 
the cross section~(\ref{eq_dcross}) is plotted in Fig.~\ref{fig:edycrossPLB} as a function of $|t|$ with $\tau=0.2$, which plays role similar to the Bjorken variable in the DIS, and also as a function of $\tau$ for a fixed value of $|t|= 0.2$~GeV$^2$. The results of \cite{Berger:2001zn} are labeled  as
``BMP2001'', where a model of the nucleon GPDs $\tilde H$ and $\tilde E$ based on the double distributions and the asymptotic pion DA for $\phi_\pi$ are used. The black solid curve shows (\ref{eq_dcross}),
and the dashed, dotted, and dash-dotted curves show the contributions of the respective terms in
the cross section. 
We see that the cross section is of the pb level, and thus its measurement will be challenging. The results also show dependences on $t$ as well as $\tau$; indeed, the cross section decreases for increasing $s$.

Recently we have updated the estimate of the corresponding cross section~\cite{Sawada:2016mao}
and obtained the results labeled as ``GK2013'' in Fig.~\ref{fig:edycrossPLB}: we have used a recent parameterization for the GPDs $\tilde H$ and $\tilde E$, determined by comparing with the HERMES data for $\pi^+$ electroproduction, as well as to the pion DA with the pre-asymptotic corrections for $\phi_\pi$. We find that the updated cross section shown by the red solid curve is enhanced compared with the previous result shown by the black solid curve. For the detail, we refer the readers to \cite{Sawada:2016mao}.

Using this updated estimate of (\ref{eq_dcross}) as an input, 
we have performed the Monte Carlo simulation and the feasibility study~\cite{Sawada:2016mao} for measuring the exclusive
pion-induced Drell-Yan process,  $\pi^- p \to \mu^+ \mu^- n$, at J-PARC. 
We are able to obtain Monte Carlo simulation signals for the dimuon mass spectra for 
the secondary pion beam with the J-PARC high-momentum beam line,
assuming a minimal extension of the E50 spectrometer for the charmed-baryon spectroscopy experiment at J-PARC, such that a dedicated muon identification ($\mu$ID) system is added at the most downstream position of the spectrometer (see Fig.~9 in \cite{Sawada:2016mao}). 
This allows a good
momentum determination of muon tracks for the forward production so that the exclusive Drell-Yan
process can be characterized via the missing-mass technique.
The Monte Carlo simulated missing-mass $M_X$ spectra
for the values of the pion beam momentum corresponding to the secondary pion beam at J-PARC
are obtained and demonstrate that 
the exclusive Drell-Yan signal is well-separated from the inclusive as well as other signals, see Fig.~14 in \cite{Sawada:2016mao}. 
It is also demonstrated~\cite{Sawada:2016mao} that the accuracy expected for the corresponding J-PARC data 
of the exclusive Drell-Yan process will allow us to 
distinguish the typical parameterizations for the GPDs used in theoretical estimate of the cross section~(\ref{eq_dcross}).
Thus, we have good results for the feasibility study of the exclusive
pion-induced Drell-Yan process,  $\pi^- p \to \mu^+ \mu^- n$, with E50 spectrometer at J-PARC.
For further detail, see the discussion in \cite{Sawada:2016mao}. 

\section{Nonfactorizable mechanism}
\label{sec:3}
We move on to soft mechanism for the exclusive Drell-Yan process, 
which does not obey the QCD factorization.
As already emphasized above, 
the lower diagram of Fig.~\ref{fig1} and the diagrams of similar type 
correspond to the LO in the QCD factorization for the exclusive Drell-Yan process. To calculate the convolution implied by these diagrams, we integrate the corresponding amplitudes over the momentum associated with the gluon propagator. When the gluon momentum becomes small compared with $\Lambda_{\rm QCD}$, such soft and nonperturbative degrees of freedom should be separated into the long-distance parts in the spirit of the QCD factorization.
Absorbing the soft nonperturbative gluon propagator into either the pion DA or the nucleon GPDs 
leads to the ``tree'' diagrams, which are obtained formally by removing the gluon propagator from the lower diagram of Fig.~\ref{fig1}
and the diagrams of similar type. 
Thus, the tree diagrams correspond to the lower order in $\alpha_s$ than the LO in the QCD factorization framework
and physically represent the ``Feynman mechanism'':
the antiquark (quark) carrying almost all pion-momentum annihilates with the quark (antiquark) carrying almost all momentum-transfer from the nucleon, 
so that produces $\gamma^*$,
while the ``wee'' parton is directly transferred between the pion and the nucleon . 
The corresponding partonic process is not ensured to be of short-distance, 
and thus this diagram is not factorizable into the short- and long-distance parts. 
Moreover, we do not have a boundary to separate the pion and the nucleon wave functions
because they are directly connected by the soft parton line, 
so that the nonperturbative function to describe the tree diagrams is also nonfactorizable between the pion and the nucleon.
The soft nonfactorizable mechanism due to the tree diagrams for the exclusive Drell-Yan process was mentioned also in Sec.~6 of \cite{Berger:2001zn}; however, even a rough estimate of it is not known so far.

It is worth noting that a similar soft nonfactorizable mechanism is known to play an important role 
in the QCD description of the pion electromagnetic form factor.
In the QCD factorization formula for the pion electromagnetic form factor, the LO contribution is 
associated with a hard-gluon exchange in the partonic subprocess and is expressed as its convolution with the two DAs, 
each of which is associated with each of the external pion legs~\cite{Lepage:1980fj}. On the other hand, we have the nonfactorizable mechanism corresponding to the tree partonic-process without gluon exchange,
where the two pions are connected by a soft paton transfered direcly between them, and it is demonstrated that this nonfactorizable mechanism plays essential role to reproduce the empirical behavior of the pion form factor in QCD calculation, especially for the moderate momentum-transfer region (see e.g., \cite{Braun:1999uj}).

Thus, the nonfactorizable mechanism due to the tree diagrams could play important roles also in the exclusive Drell-Yan process.
In general, it is a difficult task to perform a QCD calculation of such soft mechanism, to which the factorization formula is inapplicable.
To estimate those mechanisms for the exclusive Drell-Yan process in a largely model-independent way, 
we first make the external leg of the initial pion {\it off-shell}, and replace the corresponding pion wave function by the axial vector vertex, to which the pion can couple.
This procedure leads to a description using the two-point correlator ($q=q'+p'-p$, $q^2 \neq m_\pi^2$),
\begin{equation}
i\int  {d^4}x{\mkern 1mu} {e^{iq' \cdot x}}\langle n(p')|{\rm{T }}j_\mu ^5(0)j_\nu ^{{\rm{em}}}(x)|p(p)\rangle\ ,
\label{corr}
\end{equation}
corresponding to the off-forward virtual Compton amplitude with one of the electromagnetic 
currents, $j_\nu ^{{\rm{em}}} = {e_u}\bar u{\gamma _\nu }u + {e_d}\bar d{\gamma _\nu }d$, replaced by the axial vector 
current, $j_\mu ^5 = \bar d{\gamma _\mu }{\gamma _5}u$.
For deeply virtual region, $| q^2|,| q'^2 | \gg \Lambda _{{\rm{QCD}}}^2$,
the correlator (\ref{corr}) can be systematically treated by the operator product expansion (OPE).
In particular, is is straightforward to see~\cite{tanaka} that the long-distance contribution of the corresponding OPE 
can be expressed by the nucleon GPDs $\tilde H$ and $\tilde E$, which are the same GPDs as appeared in the LO QCD factorization formula~(\ref{eq_dcross}).

We may also write down the dispersion relation for (\ref{corr}) with respect to its dependence on $q^2$,
and it can be shown that the residue at the pion pole, $q^2=m_\pi^2$, 
for this dispersion relation corresponds to the exclusive Drell-Yan amplitude
associated with the on-shell pion leg. 
The corresponding residue may be determined 
from the behavior of the OPE for the off-forward deeply virtual amplitude (\ref{corr});
when we perform the OPE at the tree level, the result should give the nonfactorizable mechanism due to the tree diagrams.
For an efficient matching between the OPE and dispersion relation to determine the relevant pole residue,  
we rely on quark-hadron duality to deal with the unwanted higher resonance contributions arising in the dispersion relation.
This procedure yields the soft nonfactorizable amplitude for  $q^2=m_\pi^2 \rightarrow 0$
as~\cite{tanaka},
\begin{eqnarray}
\langle n|j_\nu ^{{\rm{em}}}|\pi^- p\rangle&=&-g_\nu ^ -\frac{2i}{{{f_\pi }}}\int_\xi ^{{x_0}} {dx}
e^{ - \frac{(x - \xi)Q'^2}{(x + \xi)M_B^2}}
\left[ {e_u}{{\tilde H}^{du}}(x,\xi ,t) \right.
\nonumber\\
&&\!\!\!\!\!\!\!\!\!\!
-
\left. {e_d}{{\tilde H}^{du}}( - x,\xi ,t)\right]
\bar u(p'){\gamma ^ + }{\gamma _5}u(p )  
 +  \cdots\ ,
\label{sna}
\end{eqnarray}
in terms of the proton GPDs, ${\tilde H^{du}}(x,\xi ,t) = {\tilde H^u}(x,\xi ,t) - {\tilde H^d}(x,\xi ,t)$,
and 
we have also the similar term associated with  
${\tilde E^{du}}(x,\xi ,t) = {\tilde E^u}(x,\xi ,t) - {\tilde E^d}(x,\xi ,t)$, 
as well as the terms arising from higher-twist
corrections to the OPE for (\ref{corr}),
in the ellipses.
Here, $x_0$ is related to the threshold parameter $q_{\rm th}^2$, from which the continuum contribution in the dispersion-relation
integral for (\ref{corr})
starts, as the approximation for the higher resonance contributions invoking quark-hadron duality. 
As a result, the integral in (\ref{sna}) appears to be performed within the DGLAP
(Dokshitzer-Gribov-Lipatov-Altarelli-Parisi) region, where the contribution due to ${\tilde E^{q}}(x,\xi ,t)$ is negligible,
see (\ref{eqn:E_{tilde}}) below.
It is also worth mentioning that the factor $g_\nu^-$ in (\ref{sna}) indicates longitudinal polarization of
the produced $\gamma^*$.

Compared with (\ref{eq_Hdu}) based on the QCD factorization, the pion DA does not appear in (\ref{sna}); instead,
we have the exponential factor, $\exp(- [(x - \xi)Q'^2]/[(x + \xi )M_B^2])$,
depending on the Borel parameter $M_B$.
The Borel parameter is an auxiliary parameter introduced through
the Borel transformation, which allows the efficient matching between the dispersion relation and the OPE for (\ref{corr})
by suppressing the higher resonance contributions in the former as well as the higher twist effects in the latter. 
Such matching procedure is characteristic of the QCD sum rule approach. 
In the present approach, the relevant nonperturbative effects arising in the ``sum rule'' are encoded in the light-cone dominated quantities, the GPDs, 
and thus 
the result (\ref{sna}) corresponds to the light-cone QCD sum rule 
for the nonfactorizable amplitude in the exclusive Drell-Yan process.
We note that the light-cone QCD sum rules have been derived for e.g., the pion electromagnetic form factor 
in \cite{Braun:1999uj}.

\begin{figure}[hbtp]
\includegraphics[width=0.45\textwidth]{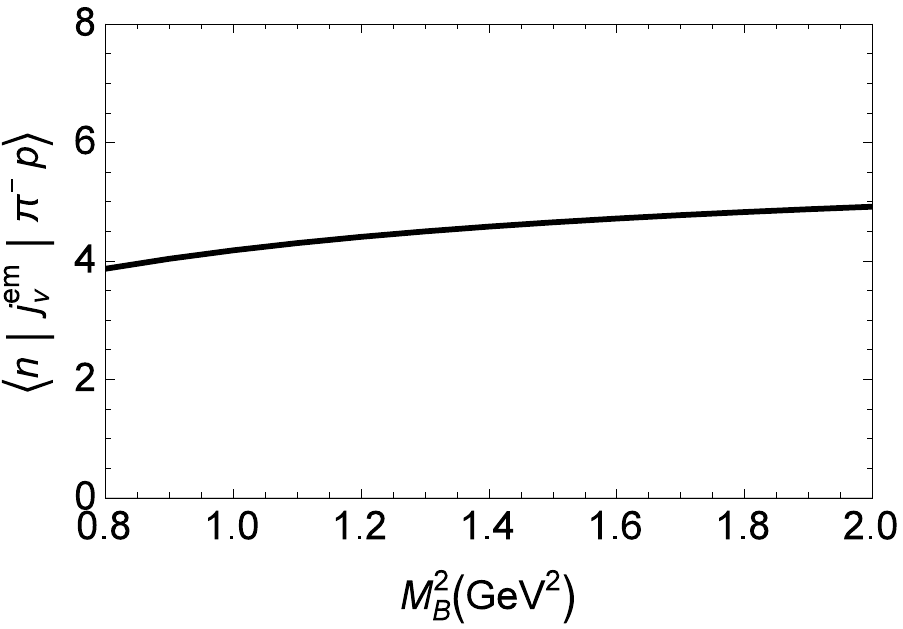}
\caption
[\protect{}] {Light-cone QCD sum rule~(\ref{sna}) for the soft nonfactorizable amplitude 
as a function of the Borel parameter squared, $M_B^2$,
with $Q'^2=5$~GeV$^2$, $|t|= 0.2$~GeV$^2$, and $\tau=0.2$, using the BMP2001 input
for the nucleon GPDs and $q_{\rm th}^2=0.7$~GeV$^2$ for the threshold parameter.
    }
\label{fig:borel}
\end{figure}

We present the behaviors of the soft nonfactorizable amplitude
from the light-cone QCD sum rule,  (\ref{sna}), using the BMP2001 parameterization for the GPDs, 
which was used in the cross section estimate shown in Fig.~\ref{fig:edycrossPLB}.
We note that the predictions using the QCD sum rules should not depend strongly on the Borel parameter, $M_B$,
introduced auxiliarily for the matching procedure.
. 
Figure~\ref{fig:borel} shows (\ref{sna}) 
as a function of  $M_B^2$
with $Q'^2=5$~GeV$^2$, $|t|= 0.2$~GeV$^2$, and $\tau=0.2$.
We obtain good stability in a relevant range for $M_B^2$.

\begin{figure}[hbtp]
\includegraphics[width=0.45\textwidth]{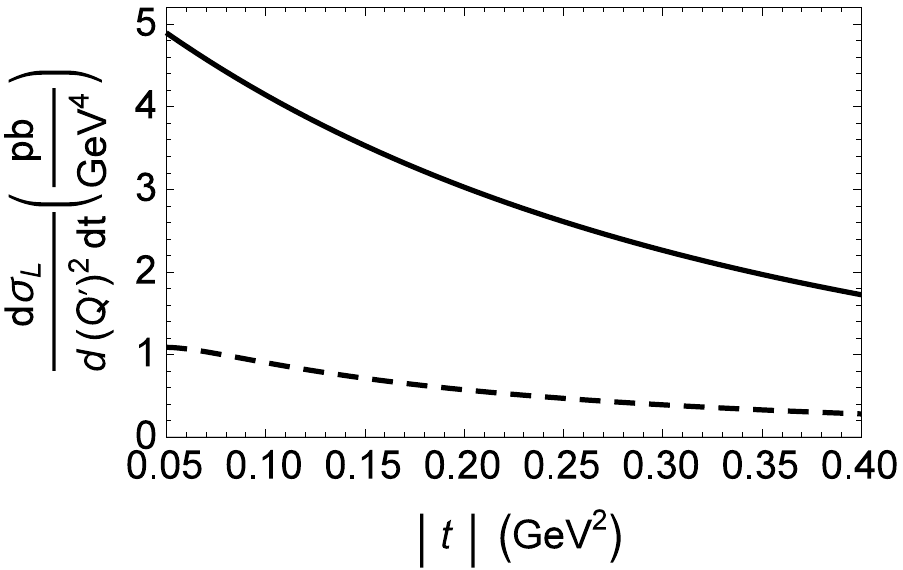}
\\ \vspace{0.7cm}
\includegraphics[width=0.45\textwidth]{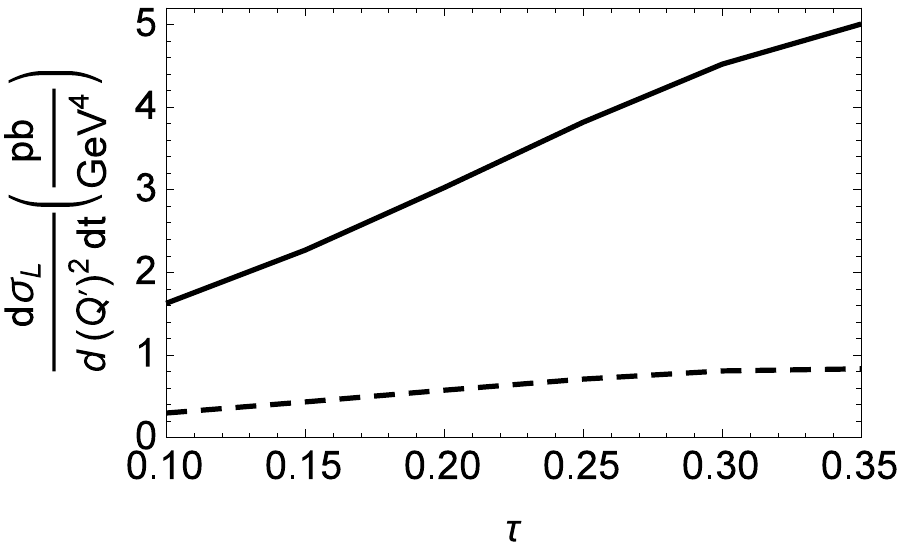}
\caption
[\protect{}] {Differential cross section
of
$\pi^- p \to \gamma^* n$ with $Q'^2=5$~GeV$^2$
as a function of $|t|$ for $\tau = 0.2$ (upper) and as a function of
  $\tau$ for $|t|= 0.2$~GeV$^2$ (lower). 
The solid curve is the prediction based on the soft nonfactorizable mechanism using the result of Fig.~\ref{fig:borel} 
from the light-cone QCD sum rule~(\ref{sna}) with the BMP2001 input
for the nucleon GPDs. The dashed curve is the prediction based on the QCD factorization mechanism with (\ref{eq_dcross}),
(\ref{eq_Hdu}) using the BMP2001 input.
    }
\label{fig:4}
\end{figure}

This result for (\ref{sna}) leads to the prediction to the cross section, $d\sigma_L/(dt dQ'^2)$,
due to the soft nonfactorizable mechanism for the exclusive pion-induced Drell-Yan process, $\pi^- p \to \gamma^* n$,
as shown by the solid curve in the upper (lower) figure of Fig.~\ref{fig:4}
as a function of $t$ ($\tau$), for the case with the same kinematics as in Fig.~\ref{fig:edycrossPLB}.
For comparison, we also plot the cross section based on the LO QCD factorization formula by the dashed curves in Fig.~\ref{fig:4}, which are same as the black solid curve in Fig.~\ref{fig:edycrossPLB}.
The soft nonfactorizable mechanism gives the cross section larger by a factor of $\sim 5$ than the that based on the QCD factorization.
The former also shows the stronger dependence on $t$ as well as $\tau$ than the latter. 
The considerable difference in size between the curves in Fig.~\ref{fig:4} reflects the different order in $\alpha_s$
associated with the relevant mechanisms, $O(\alpha_s^0)$ for the soft nonfactorizable mechanism plotted by the solid curve
and $O(\alpha_s^2)$ for the LO in the QCD factorization plotted by the dashed curve.

\begin{figure}[htbp]
\begin{center}
\centering
\includegraphics[width=0.33\textwidth]{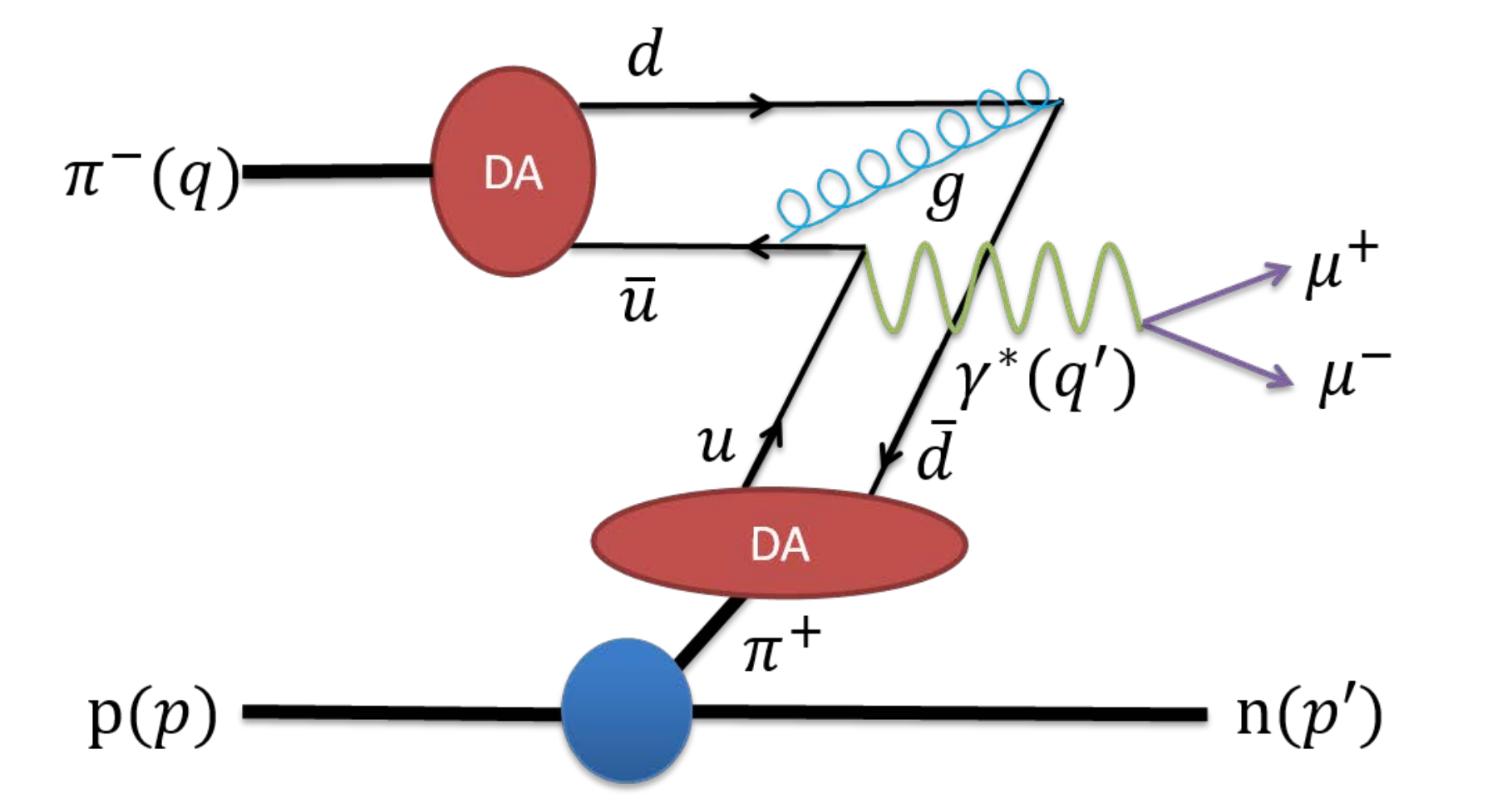}
\caption
[\protect{}]{Pion pole contribution to the distribution $\tilde{E}$.}
\label{fig:pion_pole}
\end{center}
\end{figure}

 
 We here mention the connection of the present result with the recent work by Goloskokov and Kroll (GK)~\cite{Goloskokov:2015zsa}. 
The GPD $\tilde E$ arising in $\tilde{\cal E}^{du}$, which is given by (\ref{eq_Hdu}) with the replacement $\tilde H \to \tilde E$,
is known to be dominated by the pion-pole contribution expressed as,
\begin{equation}
\tilde{E}^{u}(x,\xi,t)=-\tilde{E}^{d}(x,\xi,t)=\Theta(\xi-|x|)\frac{F(t)}{2\xi}\phi_{\pi}(x/\xi),
\label{eqn:E_{tilde}}
\end{equation} 
with
the step function $\Theta$, the
nucleon pseudoscalar form factor $F(t)$, and the pion DA $\phi_\pi$.
Thus, the substitution of this form
into the QCD factorization formula (\ref{eq_dcross})
gives rise to the mechanism of the type depicted in Fig.~\ref{fig:pion_pole}. 
In this diagram, we recognize the structure corresponding to 
the LO contribution in the QCD factorization formula for the timelike electromagnetic form factor of the pion,
\begin{equation}
F^{\rm LO}_{\pi}(Q'^2)\sim \phi_{\pi}
\otimes\alpha_s T^{(0)}_H(Q'^2) \otimes\phi_{\pi}.
\label{feltw2}
\end{equation}
as the
convolution of the (LO) partonic hard scattering amplitude, $\alpha_s
T^{(0)}_H(Q'^2)$, with the two DAs~\cite{Lepage:1980fj}.
GK~\cite{Goloskokov:2015zsa}
treated this type of contribution, separately from the factorization framework, 
as the hadronic one-particle-exchange amplitude making the replacement, $F^{\rm LO}_{\pi}(Q'^2)\rightarrow F^{\rm exp}_\pi(Q'^2)$,
where $F^{\rm exp}_\pi(Q'^2)$ denotes the experimental value of the timelike pion form factor.
It was
found that the resulting forward production cross section was enhanced by about a factor of 40, compared to the result 
within the QCD factorization formula 
corresponding to the solid curve in Fig.~\ref{fig:edycrossPLB}.

Indeed, we have,
$Q'^2 |F^{\rm exp}_\pi(Q'^2)|\simeq 0.88$~GeV$^2$, to be compared with
$Q'^2|F^{\rm LO}_{\pi}(Q'^2)|\simeq
0.15$~GeV$^2$;
for this large difference at the moderate value of $Q'^2$, 
the soft nonfactorizable for the pion electromagnetic form factor should play important role,
as already mentioned above (\ref{corr}).
This fact indicates that the huge enhancement found by GK~\cite{Goloskokov:2015zsa},
as well as the enhancement in our present result of Fig.~\ref{fig:4},
is related to the soft nonfactorizable mechanism relevant to the exclusive Drell-Yan process,
suggesting that the soft nonfactorizable mechanism could be very important at the J-PARC kinematics.
However, the contribution of the soft nonfactorizable mechanism in the present treatment 
is determined by $\tilde H$, as noted below (\ref{sna}),
while the contribution treated by GK is determined by $\tilde E$, as shown in Fig.~\ref{fig:pion_pole};
this is a consequence of the fact that the light-cone sum rule using the tree term in the OPE of (\ref{corr}) probes only the DGLAP region
for the GPDs, while the hadronic one-particle-exchange represented in Fig.~\ref{fig:pion_pole} 
probes the ERBL (Efremov-Radyushkin-Brodsky-Lepage) region.
Further study, taking into account also higher order effects,
will be needed to clarify the relation between these two approaches, and to determine quantitative roles
of the soft nonfactorizable mechanism at the J-PARC kinematics.

\section{Conclusions}
\label{sec:summary}

We have discussed the exclusive pion induced Drell-Yan process at J-PARC which allows us to probe the nucleon GPDs. We have 
presented the update of the cross section estimate for $\pi^- p \to \mu^+\mu^- n$, based on the QCD factorization formula. 
With the new estimate of the cross section, we have performed the feasibility study 
for measuring the exclusive Drell-Yan process at J-PARC and obtained good results assuming minimal extension of E50 spectrometer.
Realization of the measurement of the exclusive
pion-induced Drell-Yan process at J-PARC will 
offer the possibility
of experimentally accessing nucleon GPDs at large timelike virtuality.

We have also pointed out the soft nonfactorizable mechanism in the exclusive Drell-Yan process, the mechanism that does not obey  QCD factorization. We have shown its light-cone sum rule estimate constructed from the OPE for the off-forward deeply virtual amplitude between the axial vector current and the electromagnetic current. The results are expressed by the nonperturbative quantities, the GPDs $\tilde H$ and $\tilde E$, and the threshold parameter $q_{\rm th}^2$ through quark hadron duality. 
Our results indicate that the soft nonfactorizable mechanism gives much larger cross section than the conventional 
mechanism taken into account by the QCD factorization formula, reflecting the different order in $\alpha_s$ for those mechanisms. 

Further study of the exclusive pion-induced Drell-Yan process at J-PARC 
from theoretical as well as experimental side
will be useful for providing a new test of QCD descriptions of a novel class of hard exclusive reactions, 
for investigation of other exclusive processes at J-PARC~\cite{Kumano:2009he,Kawamura:2013iia,Chang:2015ioc}, 
and for understanding the behaviors of the GPDs and the related nonpertubative quantities~\cite{Pire:2016gut,Kawamura:2013wfa}.
This
will also clarify the interplay in the soft/hard QCD mechanisms relevant for the J-PARC processes.

\begin{acknowledgments}
I thank 
T.~Sawada, W.~C.~Chang, S.~Kumano, J.~C.~Peng, and S.~Sawada
for helpful discussions and for collaboration in \cite{Sawada:2016mao}.
I thank H.~Kawamura and P.~Kroll for useful discussions.
This work was supported in part by Grant-in-Aid for Challenging Exploratory Research No. 25610058
and Grant-in-Aid for Scientific Research (B) No. 26287040 of 
the Ministry of Education, Culture, Sports,
Science and Technology.
\end{acknowledgments}



\bibliography{apssamp}

\end{document}